\documentclass{article}
\usepackage{spconf,amsmath,graphicx}
\usepackage{algorithm2e}
\usepackage{hyperref}

\title{speech BERT embedding for improving prosody in neural TTS}
\name{Liping Chen, Yan Deng, Xi Wang, Frank K. Soong, Lei He}
\address{Microsoft China\\
{\small \tt \{lipch, yaden, xwang, frankkps, helei\}@microsoft.com}}

\begin{document}
\ninept
\maketitle
\begin{abstract}

This paper presents a speech BERT model to extract embedded prosody information in speech segments for improving the prosody of synthesized speech in neural \emph{text-to-speech} (TTS). As a pre-trained model, it can learn prosody attributes from a large amount of speech data, which can utilize more data than the original training data used by the target TTS. The embedding is extracted from the previous segment of a fixed length in the proposed BERT. The extracted embedding is then used together with the mel-spectrogram to predict the following segment in the TTS decoder. Experimental results obtained by the Transformer TTS show that the proposed BERT can extract fine-grained, segment-level prosody, which is complementary to utterance-level prosody to improve the final prosody of the TTS speech. The objective distortions measured on a single speaker TTS are reduced between the generated speech and original recordings. Subjective listening tests also show that the proposed approach is favorably preferred over the TTS without the BERT prosody embedding module, for both in-domain and out-of-domain applications. For Microsoft professional, single/multiple speakers and the LJ Speaker in the public database, subjective preference is similarly confirmed with the new BERT prosody embedding. TTS demo audio samples are in \url{https://judy44chen.github.io/TTSSpeechBERT/}.

\end{abstract}
\section{Introduction}
\label{sec:intro}
In recent years, the fast development of neural end-to-end models\cite{sotelo2017char2wav,wang2017tacotron,li2019neural,ren2019fastspeech} has boosted TTS performance. Such models aim at mapping the input text sequence to the sequence of mel-spectrograms with an encoder-decoder structure. Among them, Transformer\cite{li2019neural,ren2019fastspeech} has been proposed and widely used, referred to as Transformer TTS in this paper. Combined with neural vocoders like WaveNet\cite{oord2016wavenet}, Transformer TTS is able to generate speech waveform of high quality.

In neural TTS, prosody has long been a research topic of the community, especially in style-related tasks. In \cite{wang2018style, stanton2018predicting}, style tokens were used to model the prosody explicitly. Meanwhile, prosody can also be enriched during prosody transfer, like \cite{skerry2018towards,hsu2017unsupervised,lee2019robust,zhang2019learning}. The prosody attributes of an entire utterance or segment were extracted with a single latent variable from a reference utterance or segment which were used to control the prosody of the synthesized speech.

\begin{figure} [t]
\centering
    \includegraphics[scale=0.5]{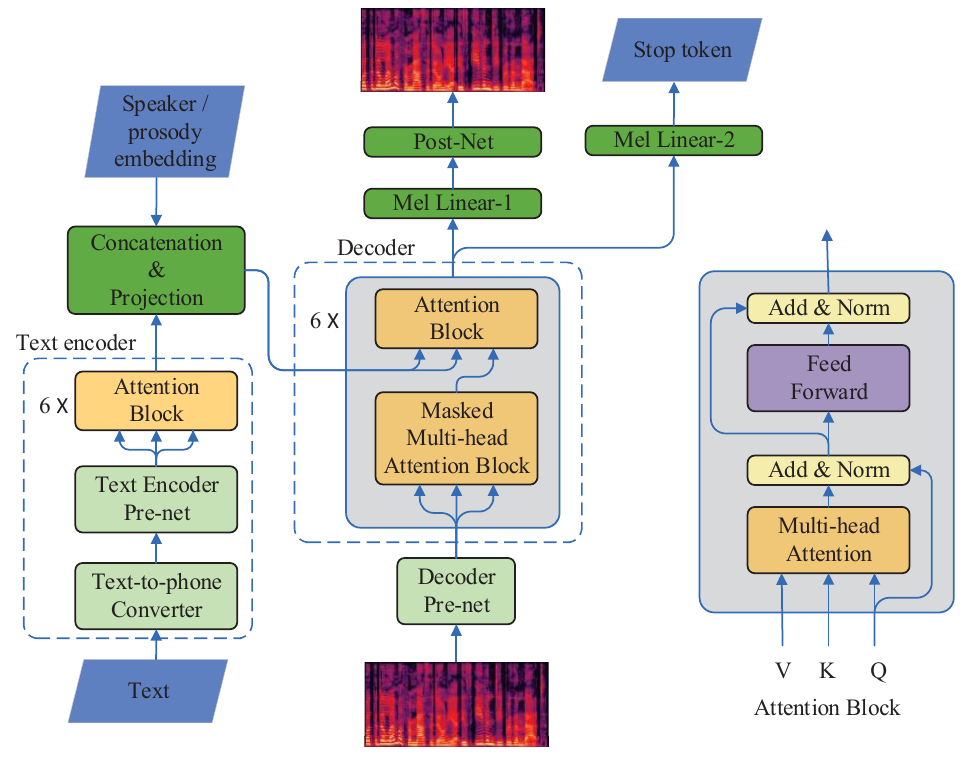} 
\caption{Transformer TTS \cite{li2019neural,ren2019fastspeech}. The attention block is specified with the module on the right side, where V, K, Q are short for value, key and query in attention, respectively.}
\label{fig. TransformerTTS} 
\end{figure}

Regarding the end-to-end sequence models, their ability in mapping phonetic and lexical information to mel-spectrogram has been acknowledged. However, with a frame-level decoder, it is still challenging for them to handle high-level expression and rich prosody variability. In this paper, we propose to improve the naturalness of the synthesized speech with the help of the syllable-level prosody attributes which can be modeled by the pre-training technique. Model pre-training is a widely used strategy in \emph{natural language processing} and \emph{automatic speech recognition} (ASR) tasks, among which \emph{Bidirectional Encoder Representations from Transformers} (BERT) has been proven successful \cite{devlin2018bert,liu2019roberta,jiang2019improving, baevski2020wav2vec}. As far as we know, this is the first work in applying such a pre-training strategy to neural TTS. 

In our work, a speech BERT model is pre-trained and used to extract embedding vectors, which are supposed to contain the prosody attributes conveyed in mel-spectrogram. Due to the Transformer-based structure of the speech BERT model, it's applied in the Transformer TTS framework for the ease of implementation. In terms of the speech BERT embedding, a speech utterance is treated as a sequence of speech segments, each of which is of about a syllable length. The embedding vectors of the previous speech segment is extracted to provide predictive information of prosody to the following segment. In transformer TTS, the embedding is adopted as the input to the decoder together with the mel-spectrogram. The advantages of such a method lie in two aspects. 1) The speech BERT model can learn the prosody attributes from a larger-scale data set than that used in Transformer TTS model, thus providing a more stable representation for the prosody attributes. Especially, for out-of-domain sentences, the prosody representation learnt from a larger data set can provide auxiliary information of how the prosody should be like. 2) The speech BERT embedding vectors are extracted on segments of syllable length, a length between a whole sentence and a single frame. As such, it handles the syllable-level prosody variation, thus providing a multi-scale variability modeling.

The rest of our paper is organized as follows. In Section \ref{sec:Transformer TTS}, we will go through the Transformer TTS model briefly. In Section \ref{sec:TTS_speech_BERT}, we will describe our speech BERT model. In Section \ref{sec:DynamicEmbedding}, the framework based on Transformer TTS which applies the speech BERT embedding dynamically is presented. We will give our experiments in Section \ref{sec:Experiments} and finally reach our conclusion in Section \ref{sec:Conclusion}.


\section{Transformer TTS}
\label{sec:Transformer TTS}
Fig. \ref{fig. TransformerTTS} shows the framework of Transformer TTS. It takes the text sequence and the mel-spectrograms of the past frames as inputs to predict the current frame. In training, the mel-spectrograms are taken from the ground-truth recordings. In inference, they are taken from the predictions of the past frames in an auto-regressive manner. So far, in neural TTS models, the speaker and prosody embeddings are input to the decoder together with the text sequence \cite{deng2018modeling,gibiansky2017deep,chen2020multispeech,9188779}. Then the speaker and the prosody style of the synthesized speech can be controlled. Overall, the text, speaker and prosody are fixed for the entire sentence in both training and inference.

\section{TTS speech BERT}
\label{sec:TTS_speech_BERT}
In this section, we will describe the TTS speech BERT, including the model structure and the speech segment masking.

\begin{figure}[t]
\centering
    \includegraphics[scale=0.4]{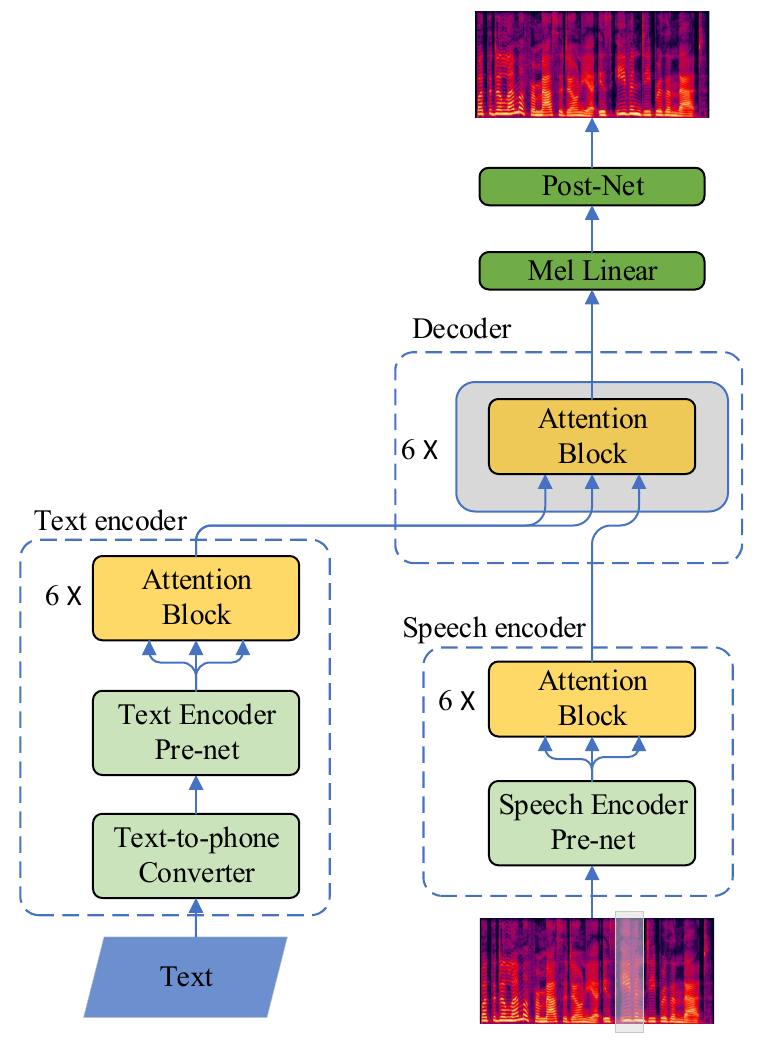} 
\caption{TTS speech BERT model. The attention block is the same as that in Fig. \ref{fig. TransformerTTS}. The segment covered by a shaded rectangular box in the input mel-spectrogram sequence is the masked speech segment.}
\label{fig. SpeechBERT} 
\end{figure}

\subsection{Model structure}
\label{sec:BERT model structure}
Our speech BERT model for TTS is derived from the Transformer TTS model as shown in Fig. \ref{fig. SpeechBERT} \footnote{This figure is different from the one used in ICASSP 2021 proceedings, pp. 6548-6552. Use this one.}. Compared with the Transformer TTS model as shown in Fig. \ref{fig. TransformerTTS}, a speech encoder block is added before the decoder. In the decoder, since BERT is bidirectional, the masked attention block is discarded as well as the modules for stop token prediction. The decoder pre-net module in Transformer TTS is renamed as speech encoder pre-net in the speech BERT encoder. Different from Transformer TTS, the speech BERT model is independent of other tags like speaker and prosody. Hence, the module of speaker/prosody embedding is removed. The model uses the reconstruction error as the training loss function, minimizing the square error between the predicted mel-spectrogram and the ground-truth.

For each training mel-spectrogram sequence, the segments corresponding to syllables are selected to be masked with a probability. The boundaries of syllables are obtained by forced alignment. The masked segments are padded with a template which will be described in the following subsection.

In the TTS speech BERT model, the content pronunciation of the masked segment is mainly reconstructed in the decoder according to the input text sequence. The speech encoder is supposed to learn other acoustic attributes of the masked spectral contour from its contexts, including prosody, speaker, etc. As such, the embedding extracted by the speech BERT encoder will be used as the representation of prosody in our work.

\subsection{Mask template}
\label{sec:BERT mask template}

\begin{algorithm}[t]
\footnotesize
	\caption{Successive DTW for acoustic segment template. $length\left(\bullet\right)$ means the number of frames in the segment.}
	\label{alg. DTW}
	\SetKwData{Left}{left}\SetKwData{This}{this}\SetKwData{Up}{up}
	\SetKwFunction{Union}{Union}\SetKwFunction{FindCompress}{FindCompress}
	\SetKwInOut{Input}{input}\SetKwInOut{Output}{output}
	\Input{phone segments ${\mathcal S}=\left\{{\mathcal S}_1,...,{\mathcal S}_N\right\}$, where ${\mathcal S}_n=\left\{{\bf s}_{n}\left(1\right),...,{\bf s}_{n}\left(I_n\right)\right\}$}
	\Output{acoustic segment template ${\bf s}$}
	\Begin{
	       \For{$n=1,...,N$}
	       {
	            \For{$i=i,...,I_n$}
	            {
	                \uIf{${\bf s}$ is NULL}
	                {
	                    ${\bf s} \leftarrow {\bf s}_{n}\left(i\right)$
	                }
	                \Else
	                {
	                    1. compute the mapping function between ${\bf s}_n\left(i\right)$ and ${\bf s}$ as $f\left(\bullet\right)=dtw\left({\bf s}_n\left(i\right),{\bf s}\right)$ \\
	                    2. \uIf{$length\left({\bf s}_n\left(i\right)\right) \ge length\left({\bf s}\right)$} 
	                    {
	                        \romannumeral1. map ${\bf s}$ to ${\bf s}_n\left(i\right)$ and get $f\left({\bf s}|{\bf s}_n\left(i\right)\right) \to \hat{{\bf s}}$ \\
	                        \romannumeral2. update the template $\bf s$ to be ${\bf s}=\frac{1}{2}\left(\hat{\bf s}+{\bf s}_n\left(i\right)\right)$
	                    }
	                    \Else{
	                        \romannumeral1. map ${\bf s}_n\left(i\right)$ to ${\bf s}$ and get $f\left({\bf s}_n\left(i\right)|{\bf s}\right) \to \hat{{\bf s}}$ \\
	                        \romannumeral2. update the template $\bf s$ to be ${\bf s}=\frac{1}{2}\left(\hat{\bf s}+{\bf s}\right)$
	                    }
	                }
	            }
	       }
	   }
\end{algorithm}

The masked segment in the input speech is replaced by a padding template, which is designed to satisfy three requirements. Firstly, it should be in acoustic space as the mel-spectrogram. Secondly, the time-varying distribution attributes among the acoustic frames can be represented. Thirdly, it should be general to pronunciation variability. To this end, the averaged phone segments from a set of speech recordings is used as the acoustic segment template. Successive \emph{dynamic time warping} (DTW) on the phone segments is applied to obtain the template as presented in Algorithm \ref{alg. DTW}.

Assume the speech recording set used for template computing to be ${\mathcal U}$. The phone set on ${\mathcal U}$ is defined as $\mathcal P=\left\{{\mathrm p}_{1},..., {\mathrm p}_{N}\right\}$ where $N$ is the number of phones. Through forced alignment with an acoustic model from ASR, we will get the boundaries of the phones in each recording. Statistically, from the perspective of phones, for the $n$-th phone ${\mathrm p}_n$, denote its segment set to be ${\mathcal S}_n=\left\{{\bf s}_{n}\left(1\right),...,{\bf s}_{n}\left(I_n\right)\right\}$ where $I_n$ is the number of segments for phone ${\mathrm p}_n$, for $n=1,...,N$. In our successive DTW, two segments are mapped to the longer one in each DTW step. Other options for setting the target segment in each DTW step may be the shorter one, any random one, etc. 

Given a template of $L$ frames, for the padding of a segment of $K$ frames, when $K \le L$, the first $L$ frames from the template can be used. When $K>L$, the template will be duplicated first.

\section{Transformer TTS with Dynamic embedding}
\label{sec:DynamicEmbedding}
\begin{figure}[t]
\centering
    \includegraphics[scale=0.35]{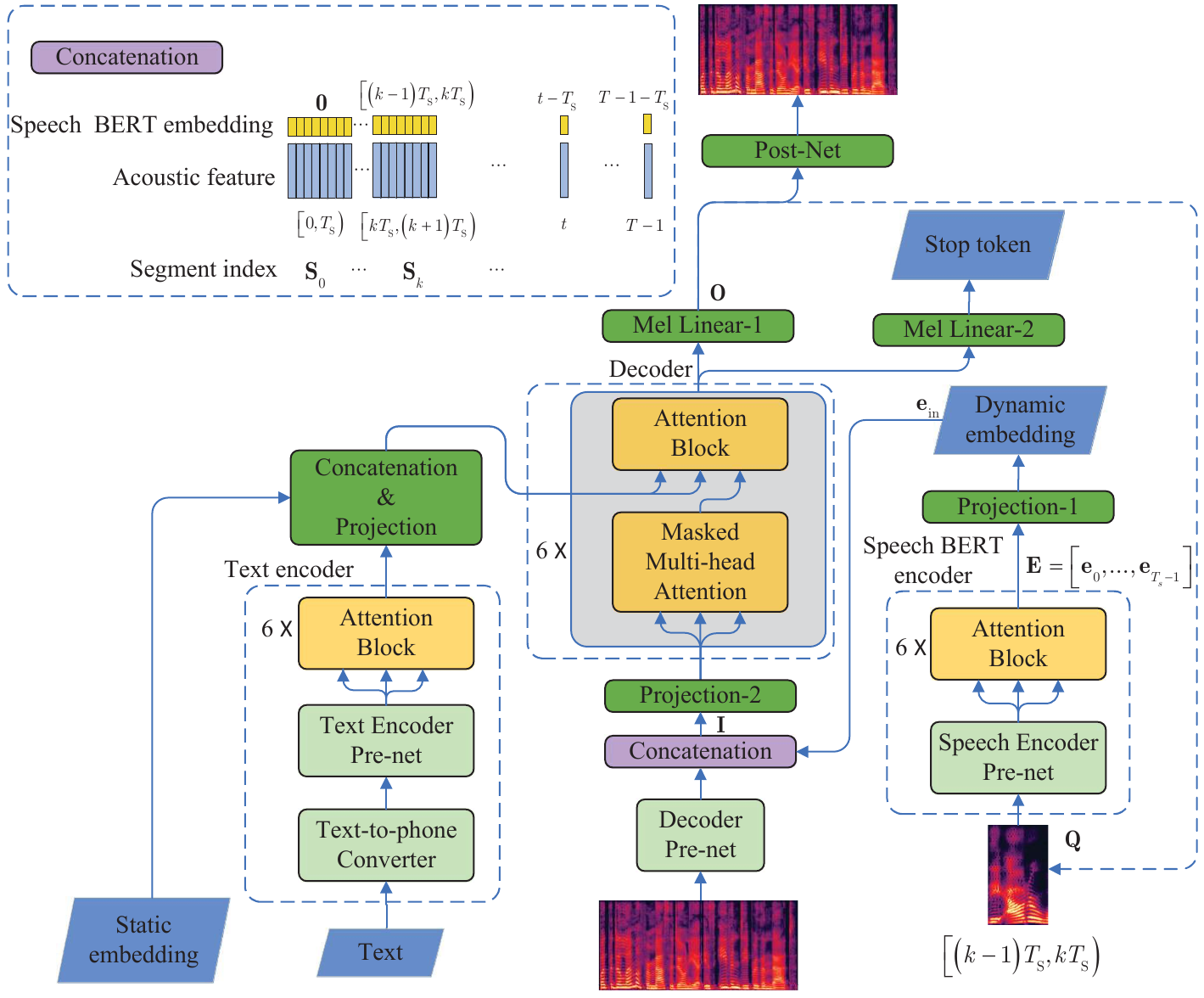} 
\caption{Transformer TTS with dynamic embedding. The attention block is the same as that in Fig. \ref{fig. TransformerTTS}. The concatenation of speech BERT embedding with the acoustic encoding is illustrated at the top left corner. The dotted arrow line from the mel linear-1 output to the speech BERT encoder infers that it's only valid in inference.}
\label{fig. dynamicTransformerTTS} 
\end{figure}

In this section, we illustrate the application of the speech BERT embedding in Transformer TTS. As presented in Fig. \ref{fig. dynamicTransformerTTS}, the speech BERT encoder is taken from the speech encoder of speech BERT model. The additional embedding on the left side with the text encoding is represented with a module named \emph{static embedding} in comparison with the \emph{dynamic embedding} on the acoustic side. Here static means to be fixed given the entire utterance. Dynamic means the embedding is extracted online given the past predicted speech.

The concatenation of the speech BERT embedding with the input mel-spectrogram vectors is shown at the top left of Fig. \ref{fig. dynamicTransformerTTS}. A mel-spectrogram sequence with $T$ frames is treated as a sequence of $K$ segments, each containing $T_{\rm S}$ frames. For the $k$-th segment, its corresponding speech BERT embedding is extracted from the $k-1$-th segment $\left(k=2,...,K\right)$. The embedding vectors are initialized to be all-zero vectors for the first segment. Before concatenation, a linear projection is applied on the speech BERT embedding. The concatenated vectors will go through a linear projection layer before feeding into decoder. During training, the speech BERT embedding is extracted from the ground-truth mel-spectrogram offline. In the auto-regressive inference, as shown by the dotted arrow line, the output vectors of the mel linear-1 layer, will be used as the approximation of the mel-spectrogram for speech BERT embedding extraction. The auto-regressive inference is illustrated in Algorithm \ref{alg. Inference}. The mel linear-1 output sequence $\bf O$ will go through the post-net for the synthesized mel-spectrogram sequence.

\begin{algorithm}[t]
\footnotesize
	\caption{Auto-regressive inference algorithm of Transformer TTS with dynamic speech BERT embedding.}
	\label{alg. Inference}
	\SetKwData{Left}{left}\SetKwData{This}{this}\SetKwData{Up}{up}
	\SetKwFunction{Union}{Union}\SetKwFunction{FindCompress}{FindCompress}
	\SetKwInOut{Input}{Input}\SetKwInOut{Output}{Output}
	\Input{input text sequence $\bf T$}
	\Output{mel linear-1 output sequence $\bf O$}
	\Begin{
	        1. Initialize speech BERT embedding matrix as ${\bf E}=zeros(T_{\rm S},d_{\rm E})$ where $d_{\rm E}$ is the dimension of speech BERT embedding\\
	        2. Initialize the mel linear vector to be ${\bf o}_0=zeros(1,d_{\rm f})$ where $d_{\rm f}$ is the dimension of the vector. Set ${\bf O}={\bf o}_0$. \\
	        3. Initialize frame index $t=0$, embedding index $i=0$ \\
	        4. Initialize the matrices ${\bf I}=[\;]$ and ${\bf Q}={\bf o}_0$.\\
	        \While{True}
	       {
	            1. Feed-forward ${\bf o}_t$ to the decoder pre-net, get ${\bf o}_{\rm in}$. \\
	            2. Feed-forward the $i$-th embedding vector ${\bf e}_i$ to projection-1 for its dynamic embedding ${\bf e}_{\rm in}$. \\
	            3. Concatenate ${\bf e}_{\rm in}$ and ${\bf o}_{\rm in}$ and append it to ${\bf I}$. \\
	            4. Feed-forward ${\bf I}$ to projection-2, decoder and mel linear-1. Get the output of frame ${\bf o}_{t+1}$. Append it to $\bf O$ and ${\bf Q}$, respectively. \\
	            5. $i=i+1$, $t=t+1$ \\
	            6. \uIf{StopToken==True}{break.}
	            7. \uIf{$mod(i,T_{\rm S})==0$}
	            {
	                \romannumeral1. Feed-forward ${\bf Q}$ to the speech BERT encoder to update ${\bf E}$. \\
	                \romannumeral2. Set ${\bf Q}=[\;]$ and $i=0$.
	            }
	       }
	       
	   }
\end{algorithm}

\section{Experiments}
\label{sec:Experiments}
Our experiments were carried out on three English data sets, an internal single speaker data set (\emph{single-spkr}) of $19.08$ hours, a public single speaker data (the LJ speaker in LJSpeech \cite{ljspeech17}) and an internal multi-speaker data set of $30$ speakers (\emph{multi-spkr}) of $247.45$ hours. For speech BERT model training, we used the public data sets including LJSpeech, VCTK \cite{yamagishi2019vctk} and LibriTTS \cite{48008}. For LibriTTS, both other and clean subsets were used. Limited by the GPU memory size, only recordings shorter than $10$ seconds ($400.66$ hours in total) were used for speech BERT training.

In our experiments, the Transformer TTS model was used as the baseline. The acoustic feature was $80$-dimensional log mel-spectrogram, extracted with a window shift $12.5$ ms. $20\%$ syllables were selected randomly for masking in each utterance during speech BERT training. The speech BERT embedding was projected to $80$ dimensions to be the dynamic embedding. The segment size $T_{\rm S}$ was $20$. A WaveNet \cite{oord2016wavenet} of 16 kHz was used as the vocoder. For single-spkr, we trained a specific WaveNet. For multi-spkr and LJ speaker, we used a WaveNet trained on multiple speakers. 

The template in speech BERT model training was computed on multi-spkr data set for two reasons. 1) The data set was recorded professionally, so the influence from noise and reverberation can be ignored. 2) The multiple speakers have various speaking styles, thus making the template more generalized. The model implementations were based on the open-source code \emph{espnet} \cite{hayashi2020espnet}. To evaluate the prosody of the two models, subjective tests were used, i.e., the \emph{mean opinion score} (MOS) and preference tests conducted by $15$ and $9$ paid native speakers, respectively.

\subsection{Objective evaluation}

Firstly, we experimented on single-spkr. In order to check the capability of the proposed method in correcting the prosody bias, we randomly selected $100$ utterances from the training set and synthesized the speech with the two models with and without speech BERT embedding. We did an objective comparison on fundamental frequency (F0), energy and duration between them as presented in Table \ref{tb. objective results}. The correlation (the larger, the better) and mean square error (the smaller, the better) on the three factors between the synthesized speech and the recordings were computed respectively. The energy and duration were computed within phones. From the comparison, we can see that, on the training set, the speech synthesized by the model with speech BERT embedding is closer to recording in terms of these prosodic variables, inferring the capability of the speech BERT in modeling and extracting the prosody attributes.

\begin{table}[h]
    \centering
    \caption{Objective comparison between without (w/o) and with (w/) speech BERT embedding. The F0, energy (E) and duration (Dur) are involved.}
    \label{tb. objective results}
    \begin{tabular}{c|c|c|c|c|c|c}
    \hline
    model & \multicolumn{3}{c|}{correlation} & \multicolumn{3}{c}{mean square error}\\
    \hline
    &F0&E&Dur&F0&E&Dur\\
    \hline
    w/o & $0.608$ &$0.922$&$0.915$&$36.057$&$17.082$& $30.67$\\
    \hline
    w/ & $0.636$ &$0.931$&$0.926$&$34.305$&$15.393$& $27.854$ \\
    \hline
    \end{tabular}
\end{table}

\begin{figure} [h]
\centering
    \includegraphics[scale=0.55]{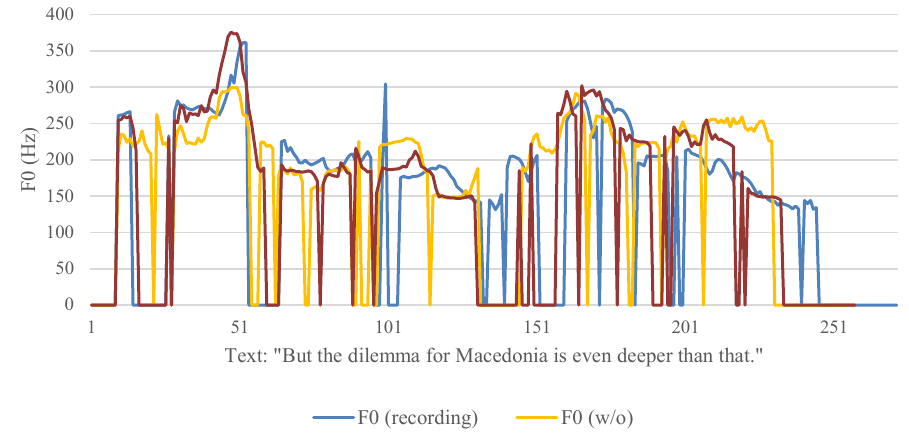} 
\caption{F0 contours on recording, synthesized speech without (w/o) and with (w/) speech BERT embedding.}
\label{fig. f0_contour} 
\end{figure}

To be specific, Fig. \ref{fig. f0_contour} shows the F0 contours generated by the recording, the synthesized speech with and without speech BERT embedding in one utterance. From the contours we can see that although there was a time shift between the synthesized speech and the recording, the F0 contour of the speech generated by the model using speech BERT embedding matches that of the recording better in trend, justifying the ability of the proposed method in correcting the bias in F0 between the synthesized speech and the recording.

\subsection{Subjective evaluation}
\begin{figure} [h]
\centering
    \includegraphics[scale=0.5]{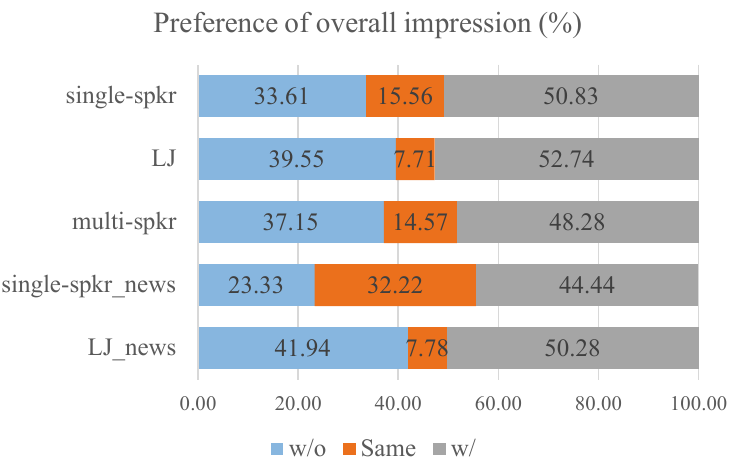} 
\caption{Results of preference tests between Transformer TTS without (w/o) and with (w/) speech BERT embedding on internal single speaker, the multiple speakers and the public LJ speaker. All the evaluations are at $p<0.01$ level.}
\label{fig. preference} 
\end{figure}

On the single-spkr, we compared the MOS scores of the recordings and the speech utterances synthesized by the Transformer TTS model with and without speech BERT embedding in Fig. \ref{tb. mos}. $40$ in-domain utterances were generated with both models. From the MOS comparison, we can see that the gap between the baseline system (w/o) and the recording was $0.15$, comparable with $0.21$ as reported in \cite{hayashi2020espnet}. In Transformer TTS, using speech BERT embedding will close the MOS gap between the synthesized speech and the recording slightly. To better compare the two models, we did a preference test between them. The result is shown in Fig. \ref{fig. preference} as \emph{single-spkr}.

\begin{table}[h]
    \centering
    \caption{MOS evaluation with $95$\% confidence intervals among recording, speech generated with Transformer TTS with (w/) and without (w/o) speech BERT embedding on single-spkr.}
    \label{tb. mos}
    \begin{tabular}{c|c|c|c}
    \hline
    &recording&w/&w/o\\
    \hline
    MOS & $4.57\pm 0.11$&$4.46 \pm 0.06$&$4.42 \pm 0.06$\\
    \hline
    \end{tabular}
\end{table}

Next, we extended the in-domain evaluations to the LJ speaker and multi-spkr. For the multi-speaker model trained on all 30 speakers, a trainable lookup table of speaker embedding is added to model speaker info. Then the speaker embedding is concatenated with the text encoder output. In the multi-spkr experiment, $10$ speakers were selected for evaluation, each with $10$ sentences, reserved from their training data. For LJ speaker, we had $50$ sentences from audio book for evaluation. The preference result between the Transformer TTS without and with speech BERT embedding is presented in Fig. \ref{fig. preference} as \emph{multi-spkr} and \emph{LJ}, respectively. From the results, we can see that on single-spkr, multi-spkr and LJ speaker, better overall impression can be obtained when speech BERT embedding is applied.

At last, we carried out the experiments in out-of-domain scenario on single-spkr and the LJ speaker. The news domain was selected which was unseen in the training sets. $40$ sentences were acquired from the website of \emph{Cable News Network} (CNN). The preference test results on the two speakers are presented in Fig. \ref{fig. preference} with \emph{single-spkr\_news} and \emph{LJ\_news} respectively. From the results, we can see that on both speakers, the model with the speech BERT embedding achieved more preference, as a result of its better prosody performance, infering us that the prosody attributes learnt from a larger-scale data set can benefit the out-of-domain speech generation.

\section{Conclusion}
\label{sec:Conclusion}

We propose a speech BERT model to extract prosody embedding for capturing salient speech prosody attributes. The speech BERT embedding  is extract dynamically, i.e., after extracted from the previous segment, and the prosody embedding is concatenated with the mel-spectrogram as input to the decoder to  predict the frames of the following segment. The objective distortions measured between the synthesized and recorded speech are reduced. The BERT can also help to reduce the bias in prosody between the predicted and the recorded speech. Subjective evaluations confirm that the TTS with the proposed speech BERT embeddings can improve the perceived prosody in both in-domain and out-of-domain scenarios. The subjective listening tests also confirm the prosody improvement is universally true for professionally recorded single/multi speakers in our proprietary clean databases and the public LJSpeech database.

\bibliographystyle{IEEEbib}
\bibliography{myref}

\end{document}